# New compact ion source design and implementation for low current applications

J. Feuchtwanger, V. Etxebarria, J. Portilla, J. Jugo, I. Badillo and I. Arredondo

*Dept. Electricidad y Electrónica. Fac. Ciencia y Tecnología, Universidad del País Vasco - UPV/EHU, 48940 Leioa, Spain*

---

**Abstract**

A new compact electron cyclotron resonance (ECR) ion source for low current applications, designed and built in-house at the University of the Basque Country (UPV/EHU) is presented. The source was designed to use as many commercially available components as possible for key tasks such as electromagnetic resonators, magnetic structure, RF power couplers and beam transport, to lower the cost and lessen reliance on specially made parts for spares. This leads to a cost-effective yet compact and fully functional design. The main design decisions are described and experiments on plasma generation and the proton beam extracted using the source are shown and discussed.

---

## 1. Introduction

Ion sources have always been important for high energy physics [1] but increasingly have gained relevance for industry, ion implantation and bio-medical environments [2]. However, most are still expensive and bulky. Their technology is a very active field of research, with clear scientific and economic interest [3].

Electron cyclotron resonance ion sources (ECRIS) have become one of the most common means for ion production in basic research and industry for a wide range of applications because of their reliability and capability to produce multiply charged ion beams from most stable elements [4]. Although much research effort has been devoted in the past to the design and development of high current, high energy sources, mostly in the context of large scientific facilities, lower current devices are of paramount importance in medical [5] [6] and industrial applications [7], as well as in low energy surface and atomic physics experiments [8], among other fields. Although many contemporary ion source designs are based on experience gathered from their preceding high energy physics examples, low current, low energy ion sources can be designed essentially from scratch so there is no need for them to be bulky or need specialty setups.

We present our new design for a compact and cost effective electron cyclotron resonance (ECR) multi-species ion source for low current applications, designed and built by the authors at the University of the Basque Country (UPV/EHU). The key factors chosen to achieve compactness and cost effectiveness are: a plasma chamber built using standard vacuum components, a permanent magnet Halbach structure to generate the ECR field, a simple RF coupler in chamber, RF power transmission over coaxial cables everywhere except for the DC break, and solid state power amplifiers.

It is difficult to compare ion sources, even among ECR ion sources, because the design parameters, the species and for what they will be used are so varied. This task is simpler if only high current proton or $H^-$ sources for large projects are compared, where the beam currents are between 30 and 100 mA, and their emittances between 0.2 and 0.3 $\pi\, mm\, mrad$ [9]. However as soon as the application is different, the spread in the parameters becomes larger with currents ranging from a few $\mu A$ to tens of $mA$ and emittances from 0.01 to almost 100 $\pi\, mm\, mrad$ [10, 11, 12, 13, 14, 15, 16, 17, 18, 19, 20].

## 2. Description of the source

The ion source is conceived for low current industrial and bio applications. Based on ECR principles, the main design parameters are summarized in Table 1. Although the table refers to $H_2$, the ion source can be operated with gases other than Hydrogen, for instance for Helium, Nitrogen or any other elemental gas for ion production. No major changes would be needed other than the use of the appropriate pressure regulator. This will be tested in the near future.

Table 1: Main design parameters of PIT30 ion source

| ECRIS Parameters | |
|---|---|
| Microwave frequency | 3 GHz |
| Microwave power | <100 W |
| Gas mass flow | <5 sccm ($H_2$) |
| Magnetic field | 110 mT |
| Extraction voltage | <20 kV |
| Beam current | <50 $\mu$A ($H^+$) |
| Beam emittance | <0.2 $\pi$ mm mrad |

A large variety of technologies had to be integrated in the source, and made to work in harmony: Mechanical, electromagnetic, vacuum, RF, high voltage and control. And while the design of all these elements has to be carried out simultaneously, because they are not independent of each other, for the structure of this paper, the criteria used in the design of each subsystem and the relation between these is detailed below in independent sections. 3GHz was chosen because solid state amplifiers for that frequency are commercially available, and the extracted current will be higher than for 2.45 GHz given that the current scales with the square to the frequency [21].

### *2.1. Mechanical design*

The main design decision taken in the design of the ion source was to use standard vacuum components in as much as possible, including the plasma chamber and the High Voltage isolators. More specifically conflat (CF) components were chosen, since the use of copper gaskets that are plastically deformed to achieve the vacuum seal has the additional advantage that it guarantees good electrical contact between the different components. In places where an even better electrical contact is required, silver plated gaskets were used, like at the plasma chamber. This type of seal is also more resistant to higher temperatures, as may be the case in the plasma chamber. The cost of the ion source was reduced by using these components, since there are several companies that manufacture them with very tight mechanical tolerances. Only a few one-of-a-kind parts were used, since specially made parts are more expensive than standard ones. Figure 1 shows a CAD rendered schematic in cross-section of what the plasma chamber built from standard components looks like. The model was made using DN 63 CF and and DN 100 CF components purchased from Vacom, Germany [22].

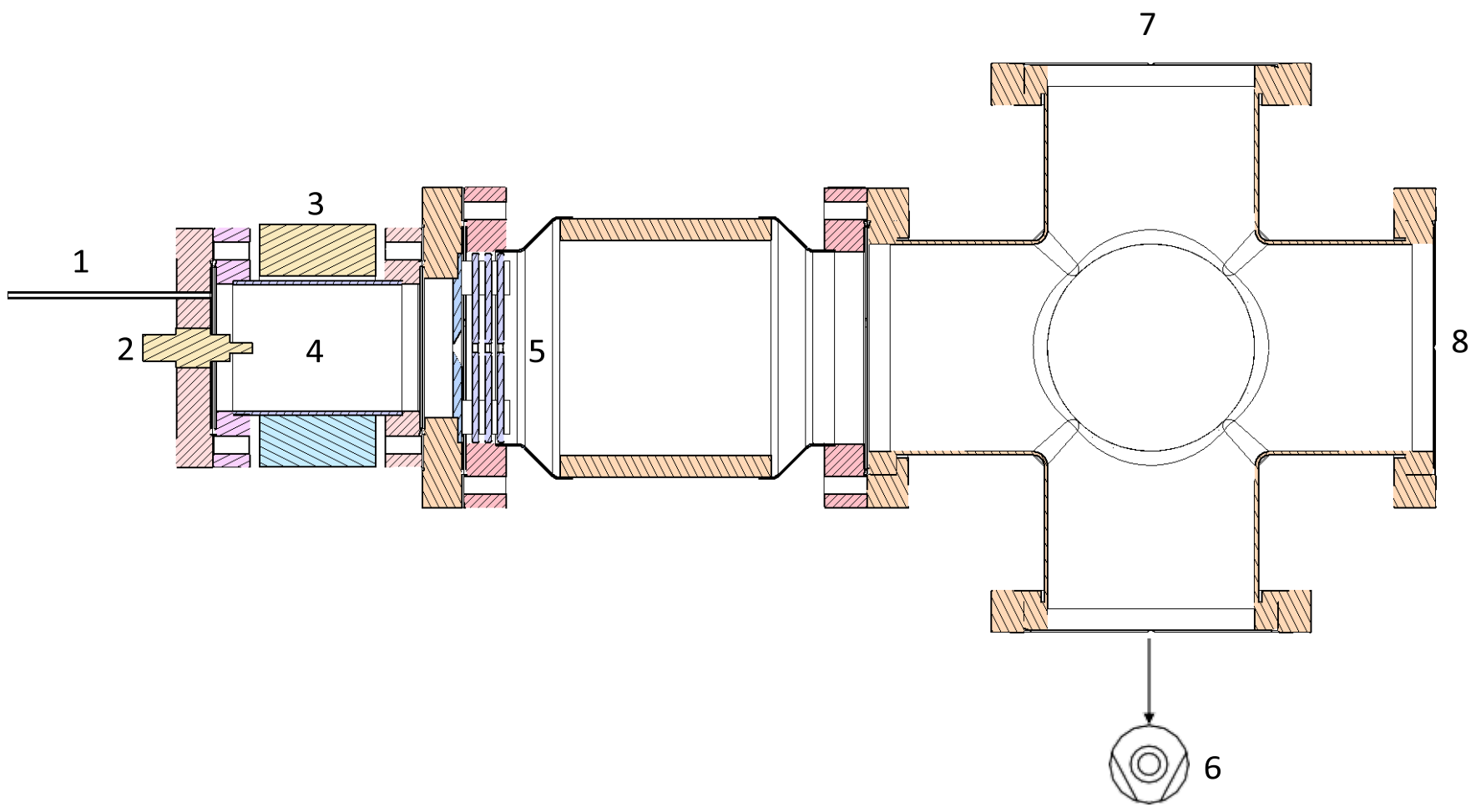

Figure 1: Cross section of a CAD drawing of the proposed plasma chamber made from standard CF components. 1) gas inlet, 2) RF port, 3) magnetic structure, 4) plasma chamber, 5) extraction electrodes, 6) Turbomolecular pump port, 7) pressure sensor 8) Faraday cup/ Scintillator screen port. The entire assembly shown is 600 mm long.

### *2.2. Electromagnetic design*

#### *2.2.1. Plasma chamber*

Since the plasma chamber was designed as a circular waveguide, for our choice of frequency the smallest commercial diameter that would act as a resonant cavity for the CF flange system was DN 63. The axial length of the chamber was calculated using analytical formulae, but because the power would have to be coupled using a loop or an antenna that would break the circular symmetry of the system, full 3D finite element simulations were used to verify the analytical calculations. Figure 2 shows a simulation of the chamber with the antenna coupler. The antenna coupler was chosen because it is simpler to make than a loop, it simply press-fits to the center pin of the rf feedthrough, it is compact and directly excites the $TE_{111}$ mode at which the cavity resonates.

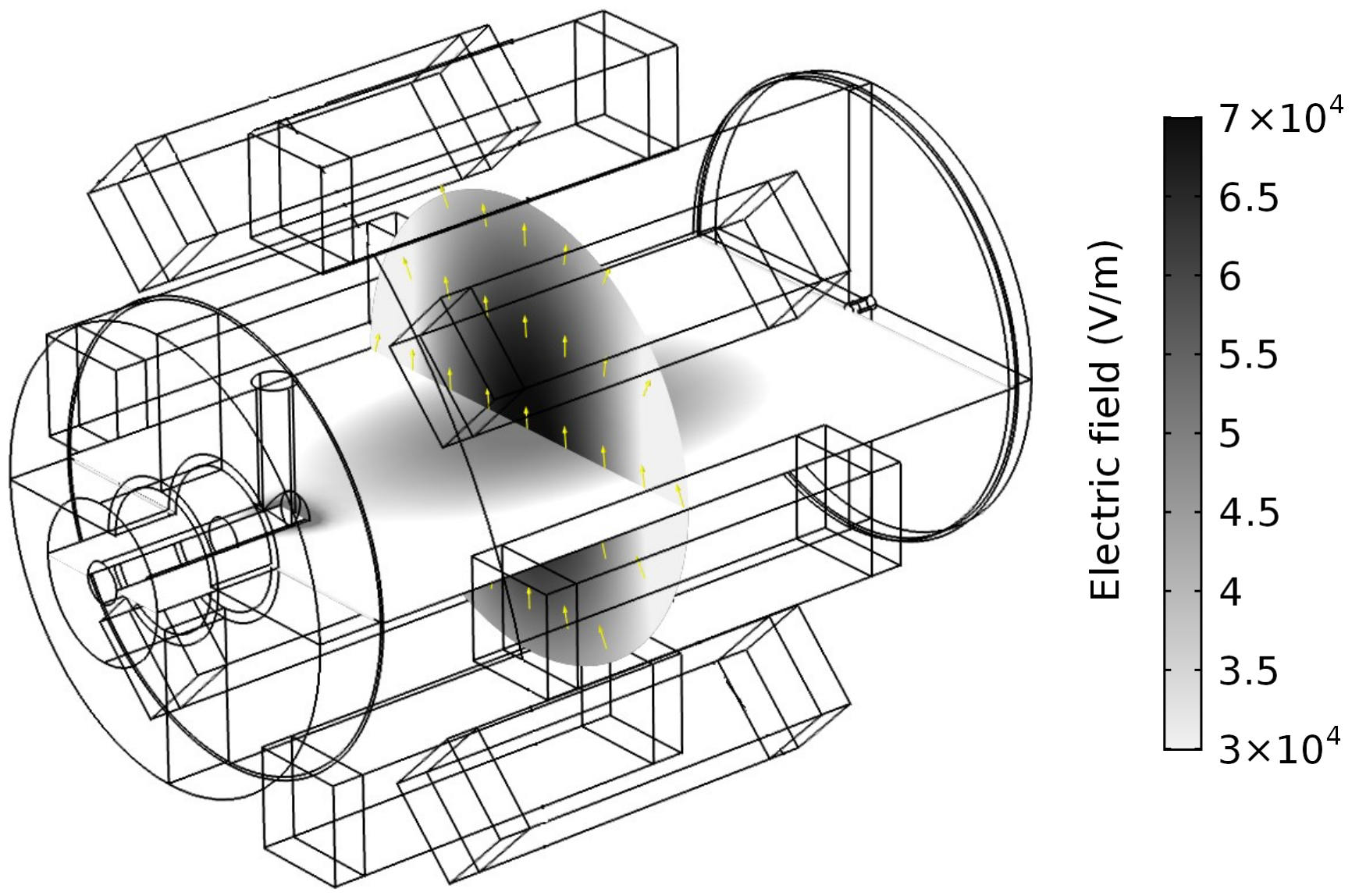

Figure 2: Preliminary simulation of the electric field in the plasma chamber generated by a 3 GHz drive.

As mentioned before, in order to reduce the cost and simplify the operation of the source, there is no electrical power supply on the high voltage end of the source. In order to generate the magnetic field needed in the chamber for the free electrons to resonate, permanent magnets were used. The intensity of the field needed was calculated from the equation for the resonant frequency of a free electron in a magnetic field ($B = 2\pi f \frac{m}{e}$ where $B$ is the magnetic inductance, $f$ is the frequency of the microwaves used, $m$ and $e$ are respectively the mass and charge of the electron). For the 3 GHz microwaves planned, this results in a field of approximately 110 mT. A Halbach type configuration made up of eight permanent magnet bars was designed to generate an axial magnetic field parallel to the plasma chamber. The diameter of the plasma chamber was chosen to be as small as possible in order to minimize the volume of the permanent magnets needed to generate the ECR field. Figure 3 shows that a field of about 110 mT in the center of the chamber is achieved. Magnetically soft iron (Armco® Pure

Iron) pieces at both ends are used to close the magnetic flux and confine it to the plasma chamber. The soft Iron piece at the beam extraction end of the chamber also doubles as the first electrode in the extraction system and helps to reduce the magnetic field in the extraction region of the source.

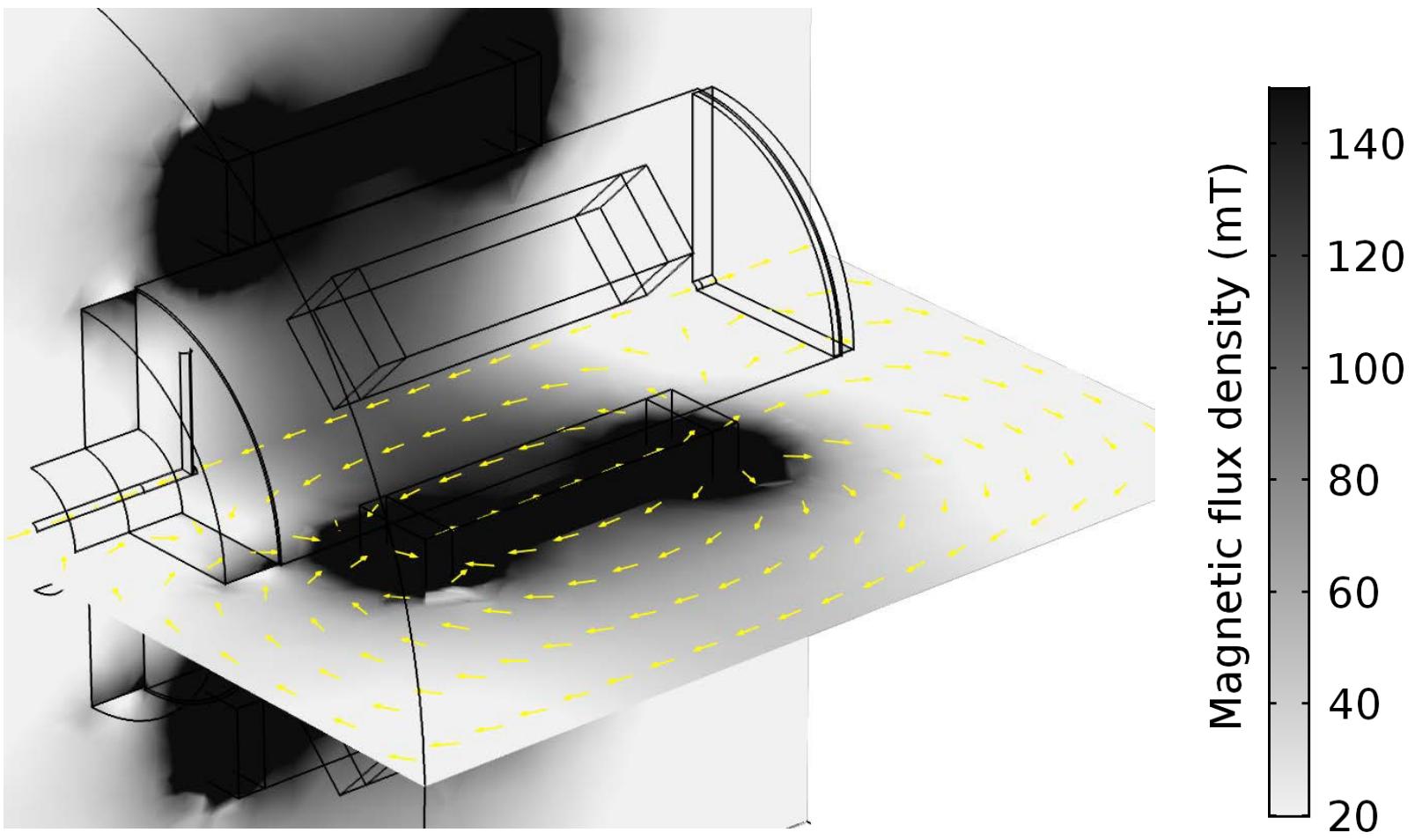

Figure 3: Simulation of the magnetic field in the plasma chamber generated by an arrangement of 8 permanent magnet bars evenly spaced around the circumference of the chamber and two soft iron pieces to help close the magnetic flux. One of the pieces also doubles as the first electrode for the extraction system

In Figure 4 it is also shown a photograph of the plasma chamber together with the Halbach permanent magnet array, which has proved to be efficient for Hydrogen plasma production [23].

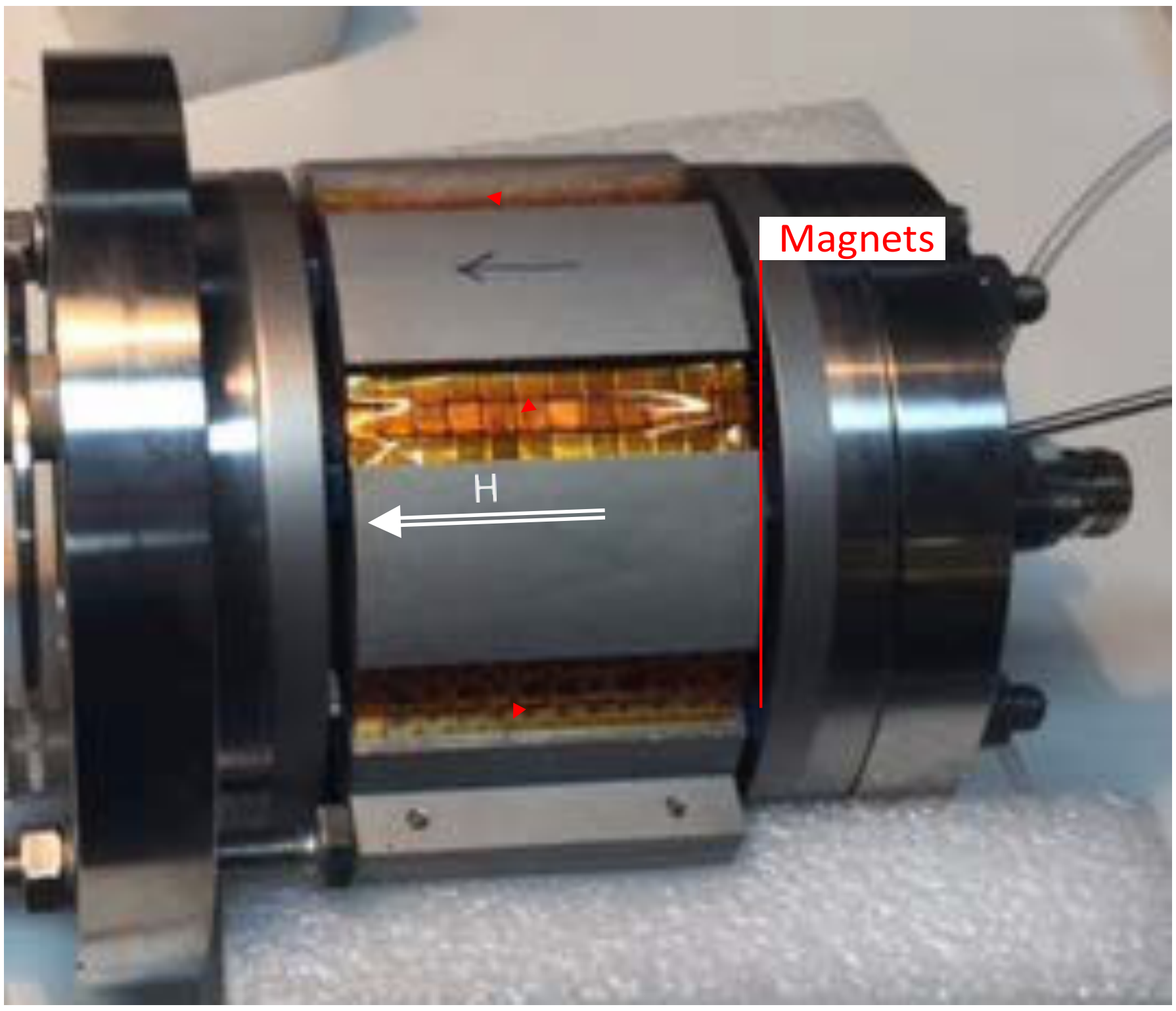

Figure 4: *Photograph of the Halbach array mounted on the plasma chamber.*

### *2.2.2. Beam extraction and optics*

Beam extraction is achieved through the use of a tetrode system. The first electrode as mentioned before also doubles as part of the magnetic system of the plasma chamber, and it has a 5 mm diameter extraction aperture with a chamfer at the pierce angle. This first electrode along with the plasma chamber are at the extraction voltage (30 kV maximum). This is followed by a set of three electrodes that are configured to work as an Einzel lens [24]. The second and fourth electrodes are connected to ground and the third electrode (central electrode in the Einzel lens) is connected to a high voltage supply, which can be varied to achieve the desired beam focus. The effect of varying this voltage is demonstrated in the results section. The inset of figure 11 shows a simulation of the extraction system using SimIon®, and shows the numbering of the electrodes.

### *2.2.3. DC break*

In order to supply the plasma chamber with RF power a DC break is used. It is made from two coaxial N to WG 284 waveguide transitions facing each other and separated by a narrow nylon separator with a central gap. The nylon piece serves both as an electrical insulator for DC voltages and as the mechanical clamp that holds the two waveguides together. A cross section of the DC break's construction can be seen in Figure 5. The design was tested in DC by connecting one waveguide to ground and the other one to 30 kV. No current was registered on the power supply during the test. The setup was then also tested at the chosen frequency of 3 GHz to verify the correct transmission of RF power. Fig 6 shows the measured RF scattering parameters $S11$ and $S21$ for the DC break as a function of frequency, where as it can be seen the RF losses at 3 GHz are only -14.8 dB. The losses during operation result in less than a $5^{\circ}$ C increase in the temperature of the DC break when the steady state is reached.

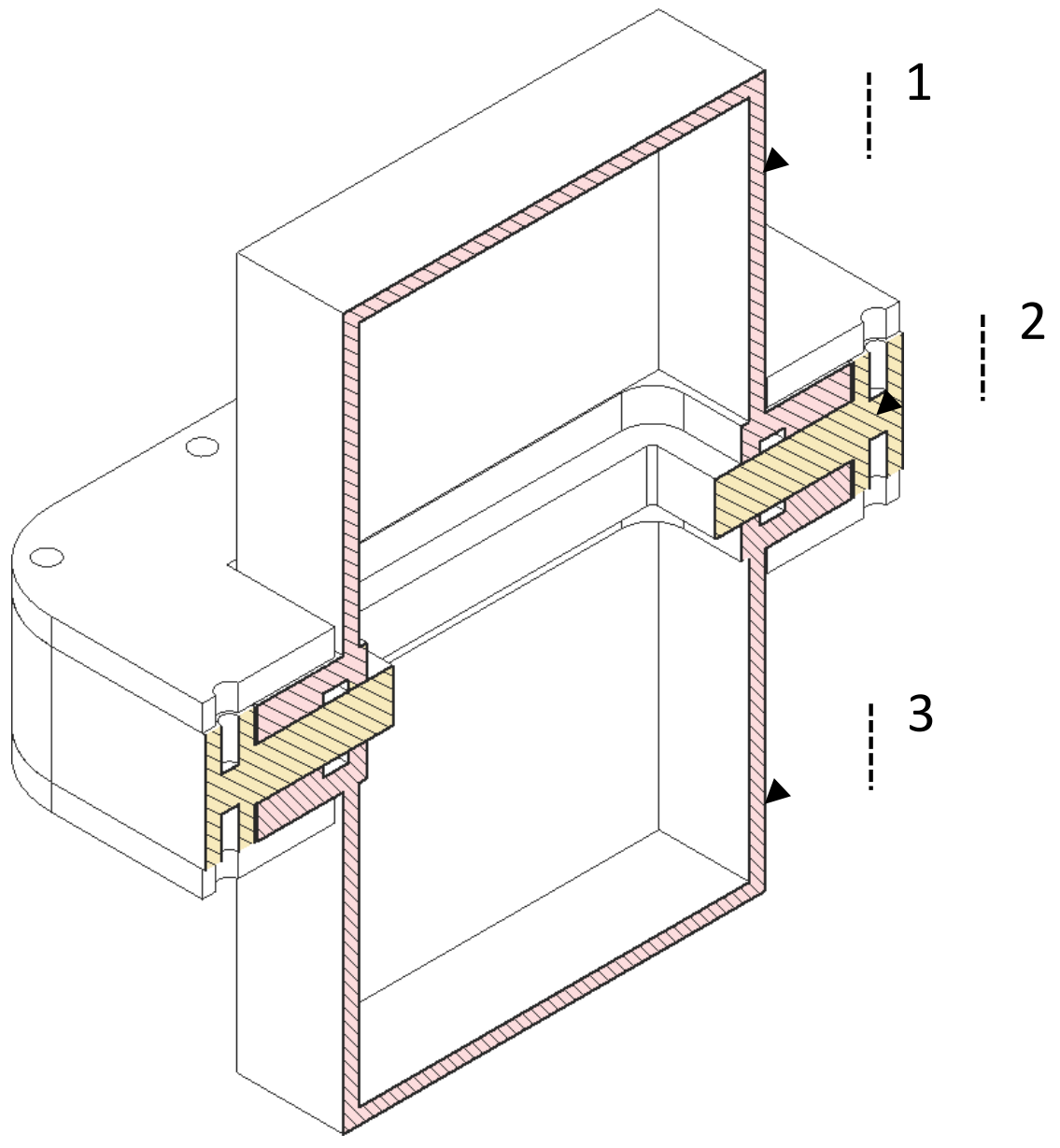

Figure 5: *Cross-section of the DC break, 1 and 3 are standard N to WG 284 transitions, 2 is the custom made nylon separator.*

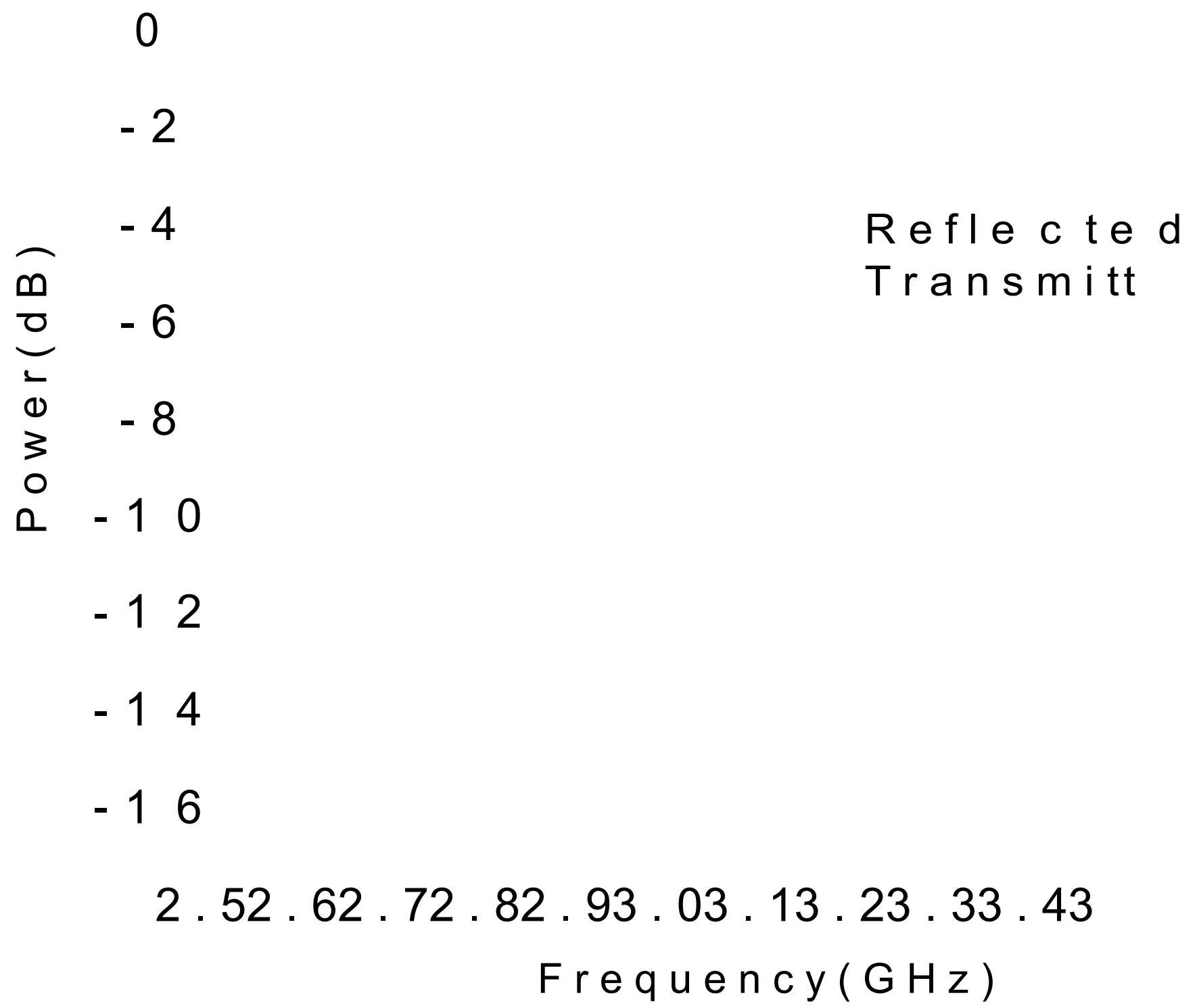

Figure 6: *Measured RF scattering parameters in dB (S11)(adaptation) and (S21) (loss) as a function of frequency for the DC break centered around 3 GHz.*

### *2.3. Vacuum system*

The vacuum pumps for the system were dimensioned by simulating the vacuum system using *Molflow*+ [25]. The system is composed of a turbomolecular pump (Pfeiffer HiPace 400) and a dry roughing pump (Adixen/Pfeiffer ACP 15) to generate the vacuum, together with a gas mass flow controller (Omega FMA-2615A) to supply the gas to be ionized. The mass flow controller provides quantity of gas independently of variations in ambient temperature. The base pressure of the system reaches $5 \times 10^{-8}$ mbar and during operation is in the $10^{-4}$ mbar range with gas flowing between 2 and 4 sccm of Hydrogen.

*2.4. Radio frequency power supply chain*

This comprises all the elements required to generate, adapt and measure the RF power used to generate the ECR plasma. A schematic of the components that make up the RF power supply chain can be seen in Figure 7. A low power signal (0 dBm max) is generated using a function generator (Agilent E 4432b) and then amplified using a GaN solid state amplifier developed for instrumentation and radiocommunication applications (MCS model AMP-2G-6G-50dBm-R) that can output up to 100 W in continuous wave mode. The output of the generator is connected to a circulator (Konnect RF, P/N: KCI 177237) and a load to protect the amplifier in case too much power is reflected. The RF power then goes through a dual directional coupler into a tree stub tuner (Maury microwave model 1878B), and then through a second dual directional coupler. The power then goes into the DC break described before and finally into the plasma chamber. The three stub tuner is used to match the impedance of the plasma chamber to the rest of the RF system and maximize power transmission to the plasma. The first directional coupler is used to measure the output of the amplifier while the second one measures the power reflected by the chamber. All the connections between the elements mentioned before are done using coaxial cables with N-type connectors.

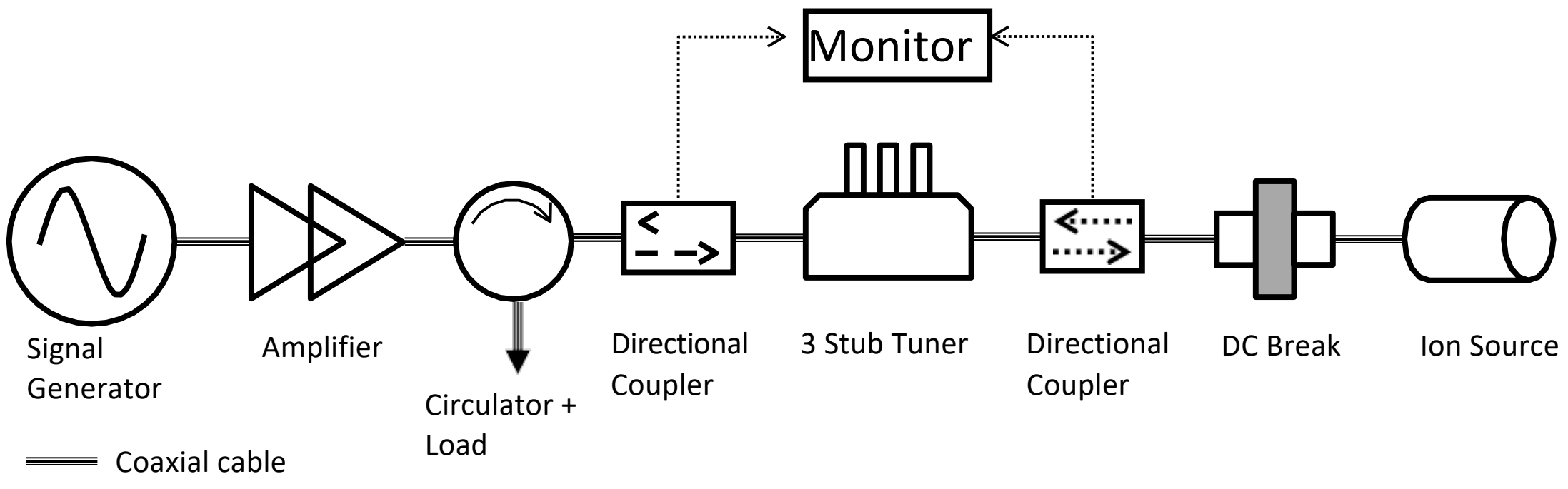

Figure 7: Schematic of the elements that make up the RF Chain used to generate the plasma in the ion source. The monitoring system takes care of all the measurements from the directional couplers.

### *2.5. High voltage system*

The extraction and acceleration of the charged particles generated in the plasma is achieved by raising the plasma chamber to a high voltage (up to 30 kV) relative to ground using an adjustable power supply (Technix SR30-P-150-OP). The plasma chamber is mechanically supported using commercial isolators rated to 30 kV normally used for medium voltage power lines. The chamber is electrically separated form the rest of the vacuum system using an alumina insulator tube brazed to CF flanges available commercially and rated to 60 kV. Gas is supplied to the chamber through a 3 mm polyamide tube that is electrically insulating. Beam focusing is achieved by supplying the 3rd electrode in the tetrode electrode described before with a second programable power supply (Stanford Research Systems PS350/5000V-25W) that can output up to ±5 kV. Both power supplies were chosen so that they can maintain a current higher than the maximum beam current in their whole operating ranges. 30 kV was chosen because it is the highest voltage that fits the Spanish legal definition of a mid tension installation.

For safety reasons all metallic components in the system (with the exception of the plasma chamber and magnet system) are connected to ground. Care was taken to avoid in as much as possible ground loops and to ensure that all connections had the highest possible conductivity [26]. The plasma chamber is enclosed in a Faraday cage made from aluminum perforated plates connected to ground so that the chamber is well ventilated because it relies on natural convection for this end. Apart from EM insulation of the source, the main purpose of the Faraday cage is to act as a physical barrier between the chamber and any living creature during operation of the source.

### *2.6. Control and monitoring*

In provision of future additions to the source, a National Instruments PXI (PXIe-1082) chassis with a real time controller (PXIe-8108) with a front end programmed in LabVIEW is used to monitor and control all the signals in the system. For future applications this configuration has the advantage that it can handle all the timing signals needed to synchronize RF-based accelerating elements and diagnostics. The power supplies, the amperemeter and the RF function generator are controlled via their IEEE 488 interfaces from the LabVIEW program. All analog signals are handled in the PXI using a PXIe-6259 DAQ card in the PXI chassis.

## 3. Beam measurements

Once the ion source was fully assembled and the individual components were tested separately, beam extraction experiments were carried out in order to validate the operability of the complete ion source. Figure 8 shows a photograph of the ion source as it stands, and where most of the elements described in the previous sections can be seen. At this point due to regulatory limitations 10 keV could not be surpassed. Three types of experiments were carried out in order to characterize the extracted beam. First, in order to verify the beam alignment, a P43 phosphorescent screen was used to look at both the position and the size of the beam spot. The misalignment of the beam was corrected by adjusting a vacuum port aligner placed between the alumina isolator and the pumping vessel. The total distance between the last electrode and the plane where all the measurements were carried out was approximately 600 mm. In Figure 9 we see the beam spot, where the brighter and darker spots on the left and top of the spot are due to prior damage to the P43 coating. In any case, the spot can be estimated to be less than 20 mm in diameter.

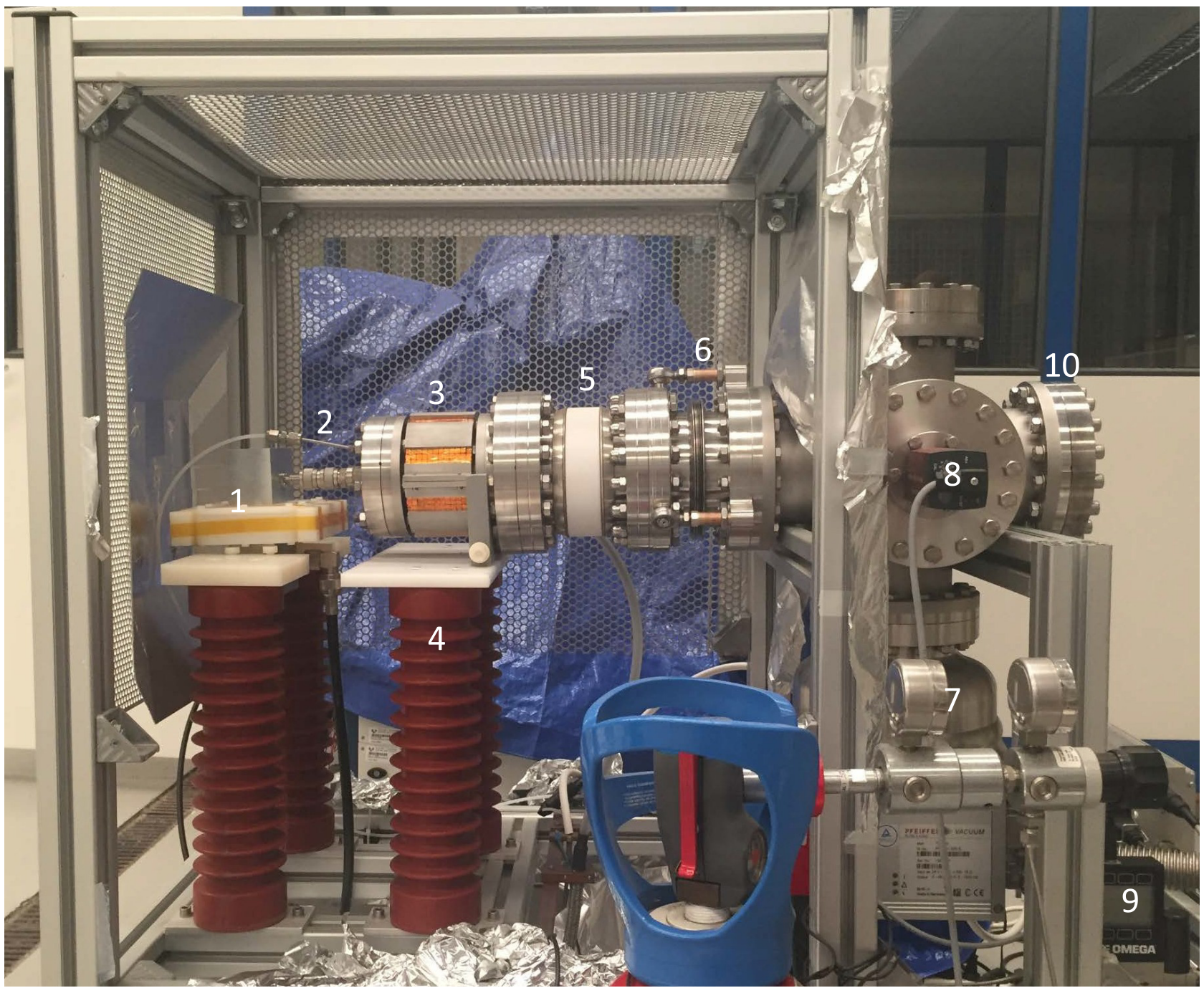

Figure 8: Photograph of the ion source as tested with one of the protective panels removed. 1) DC break, 2) Gas inlet tube, 3) Permanent magnet structure around plasma chamber, 4) High voltage stand-offs 5) Alumina isolator, 6) Port aligner, 7) Turbomolecular pump, 8) Pressure sensor, 9) Mass flow controller, 10) P43 scintillator screen.

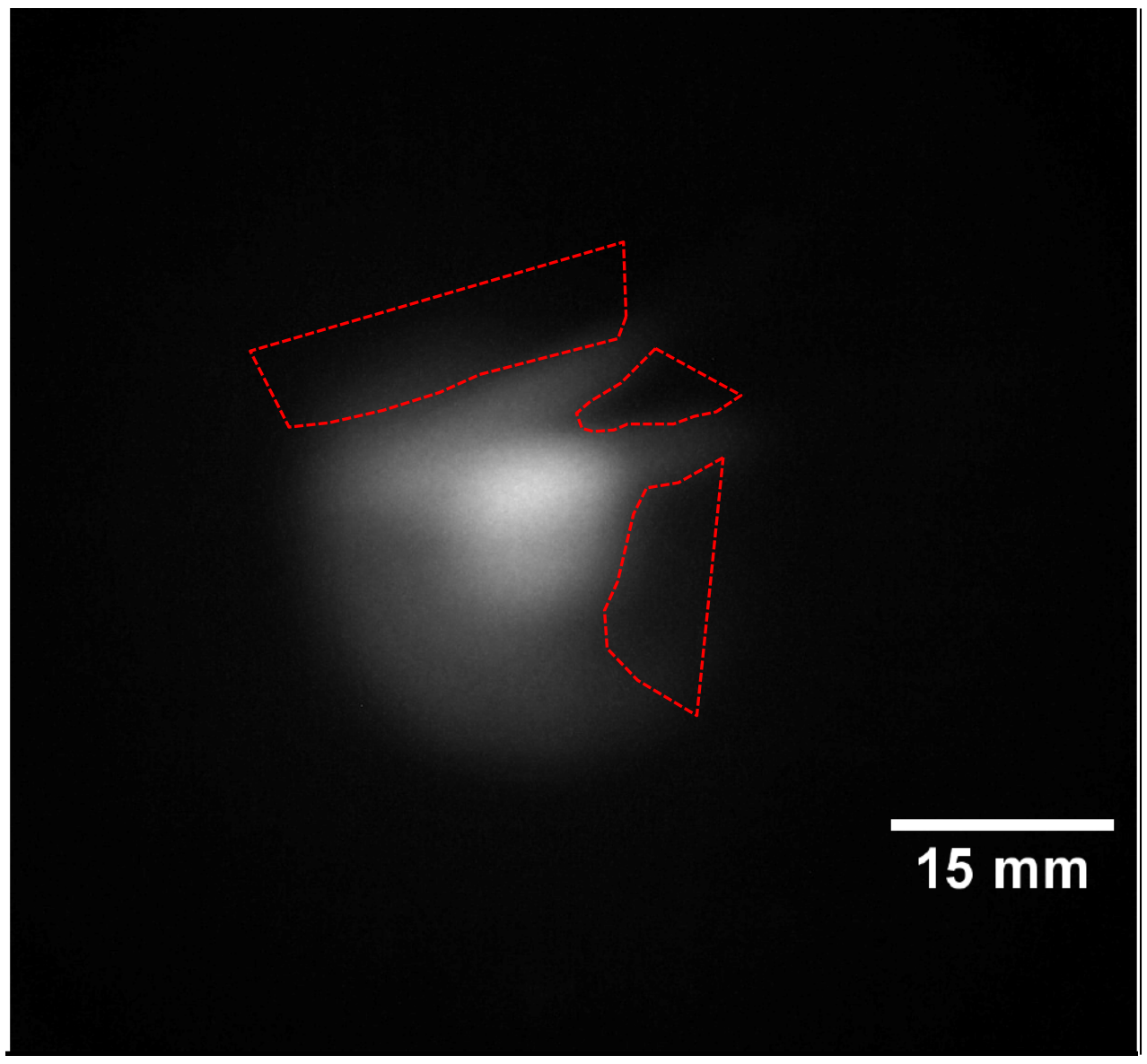

Figure 9: Image of the 6 keV beam spot on the phosphorescent screen when focused to the smallest size achievable (3.8 kV in electrode 3). The areas enclosed by the red dotted lines are the previously damaged areas.

If the intensity as a function of position of the image is analyzed, we can see that the intensity follows a gaussian distribution. This is shown in Figure 10, where the measured data (semi-transparent) are laid over the 2D gaussian that best fits the data. Despite the unevenness in the image due to the damaged screen a very reasonable fit is achieved ($R^2$=0.93). The portions of the image that correspond to the damaged sections of the screen have been marked by a red dashed line that surrounds the perimeter of the damaged area.

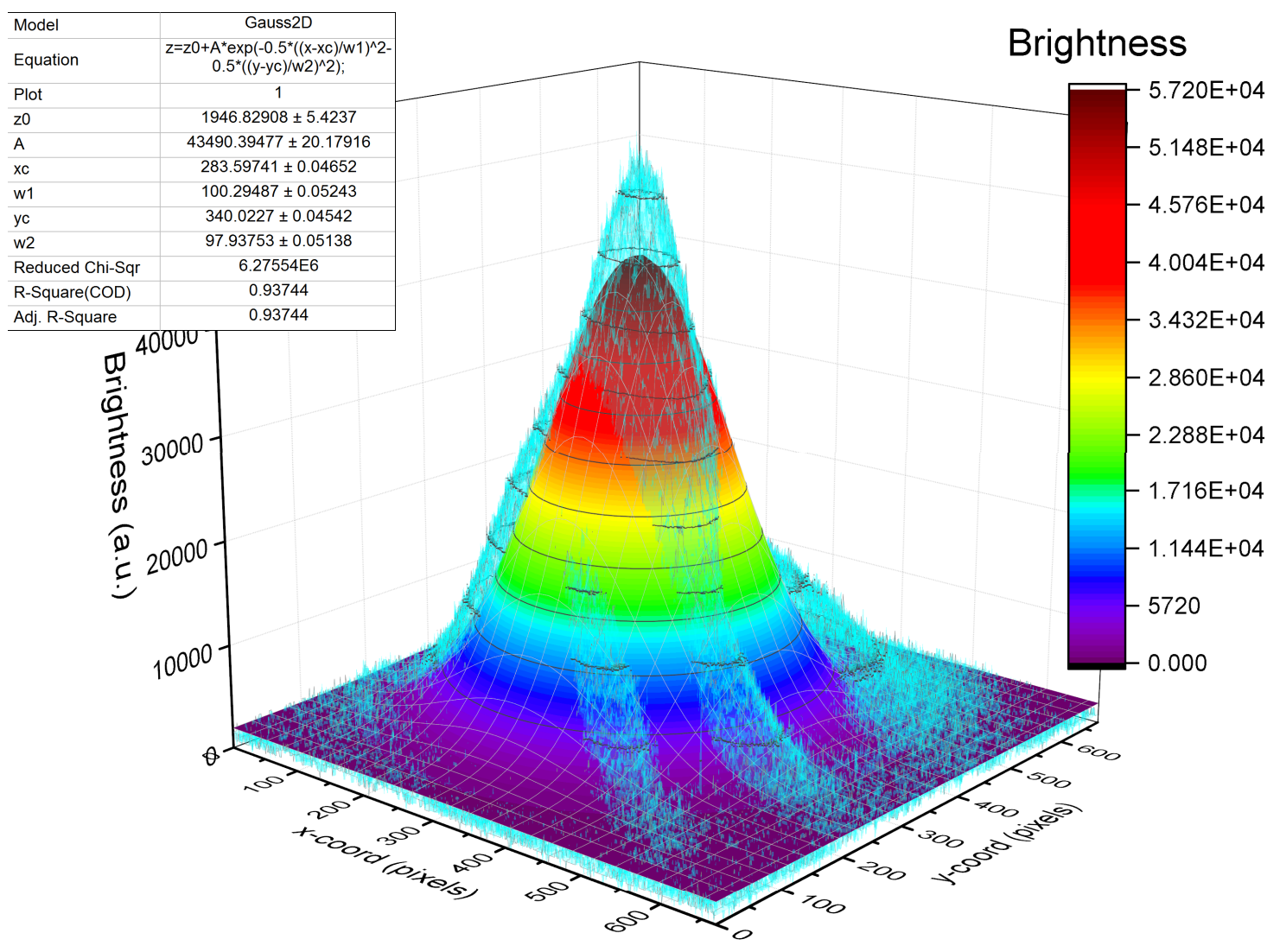

Figure 10: Image intensity for a 6 keV beam spot on the phosphorescent screen when focused to the smallest size achievable (3.8 kV in the central electrode of the Einzel lens).

Second, with the beam centered on the furthermost port down beam, the phosphorescent screen was replaced with a Faraday cup (Kimball Physics FC-73) connected to an amperemeter (Keithley Model 6430). A fixed extraction voltage was set and the Einzel lens voltage (electrode 3) was varied, as is seen in Figure 11. At first with increasing voltage an increased current was measured in the Faraday cup until a maximum was reached. Then, if the voltage was further increased, the measured current decreased. This is consistent with what was seen on the phosphorescent screen: the highest current and the smallest spot were observed at the same Einzel lens voltage for the same extraction potential. It should be noted that if both voltages were maintained, but the plasma was extinguished by turning off the RF power, no current was measured on the Faraday cup, and no image was longer visible on the screen, as expected.

With the data from Figure 11, the spot size (obtained from Figure 9), the diameter of the Faraday cup (,0 5 mm ) and the fit to a Gaussian peak from Figure 10 we can calculate the total current in the beam spot. We get that the current in the spot is approximately 6.25 times greater than what the Faraday cup captures, resulting in a total current of 4.3 $\mu$A at 6 keV.

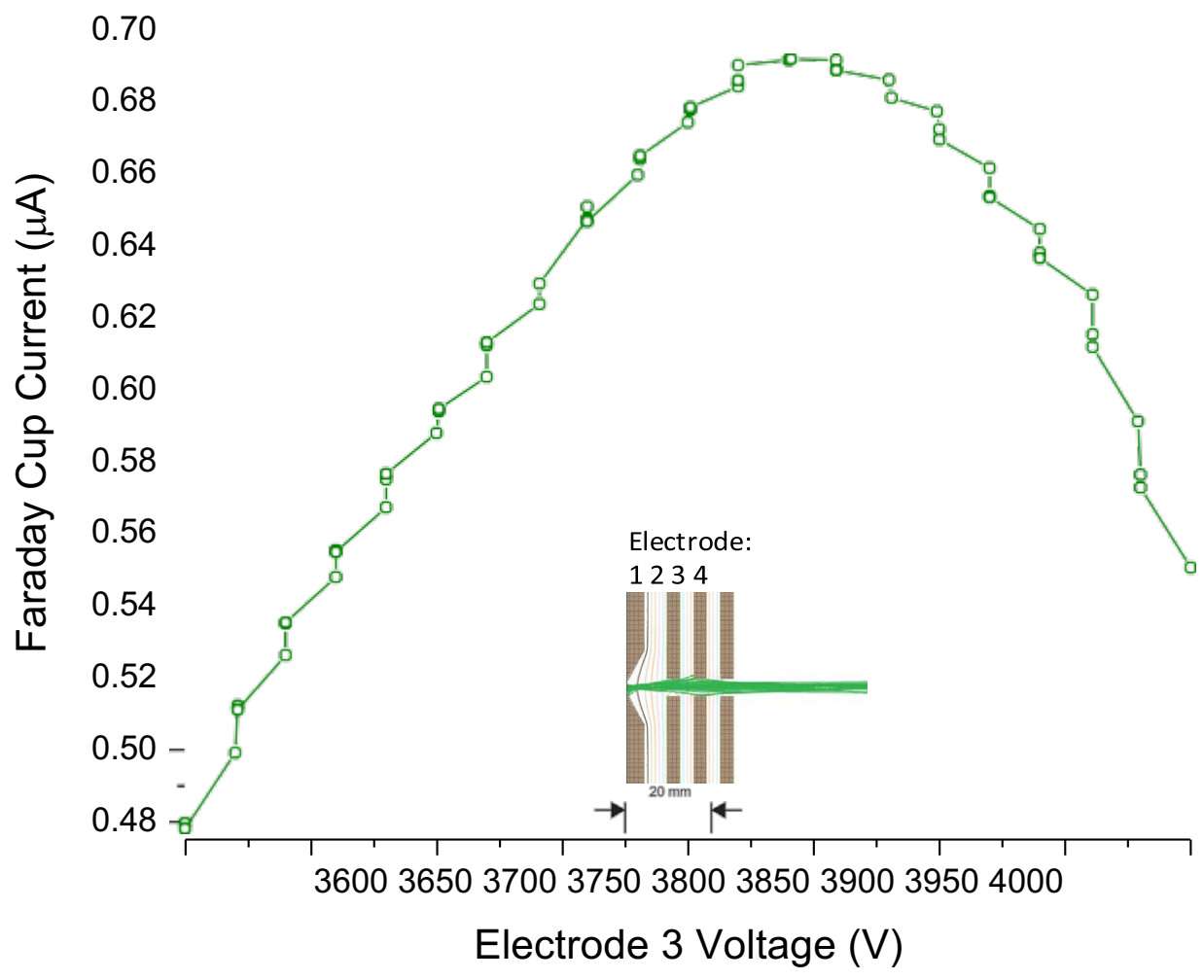

Figure 11: Beam current measured at the Faraday cup for a 6 keV beam as a function of the voltage applied to the central electrode of the Einzel lens.

Third, in order to evaluate the emittance of the source at the previous conditions, a pepperpot was installed. It consisted of a plate with a rectangular array of 0.45 mm holes spaced 1.27 mm between centers placed 37 mm from the scintillating screen. Following the procedure outlined by Zhang [28] the beam emittance was calculated, and the highest value was obtained for the horizontal direction (0.065 $\pi$mm mrad). Figure 12 shows the plots of divergence vs. position measured for the beam. The result is a narrow ellipse whose tilt indicates that, as should be expected, the beam is divergent. The emittance values obtained are very reasonable and comparable to other sources.

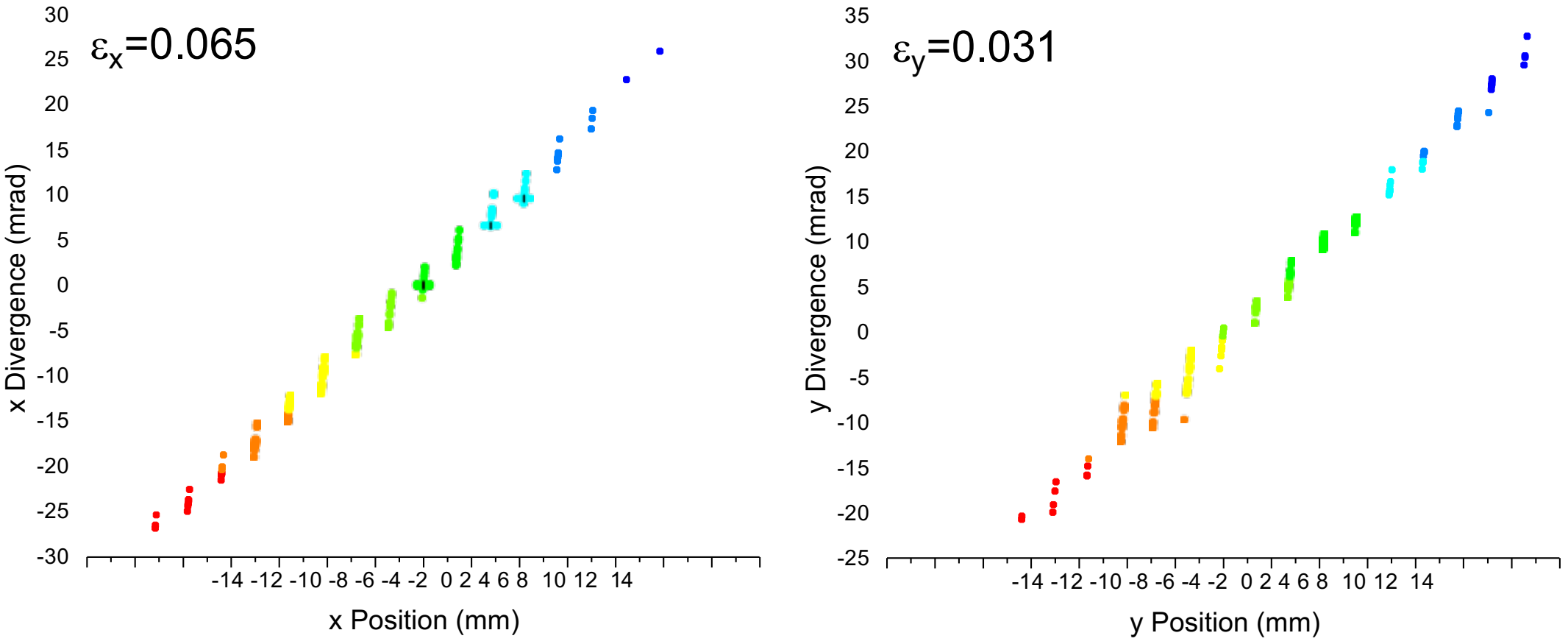

Figure 12: Horizontal and vertical divergence vs. position for a 6 keV beam, with 3.8 kV in the central electrode of the Einzel lens, obtained from the pepperpot images.

## 4. Conclusions

A newly designed compact electron cyclotron resonance (ECR) multi-species ion source for low current applications has been built and tested. As a key ingredient of the design, we have shown that using mostly off-the-shelf components is a viable way to construct a low cost yet perfectly functional off resonance ECR ion source. By keeping the RF power low (≤100 W) air cooling has been proved to be sufficient to keep the temperature of the permanent magnet assembly below the demagnetization temperature of 80°C for the N50 FeNdB magnets used, leading to a simplified source design that doesn't require any ancillary supplies on the high voltage end. The beam generated has a low emittance and a small spot size, suitable for a number of applications where low currents are needed, including biomedical and industrial uses. Because of its compactness, simplified magnetic structure, low power requirements, efficiency, quality of beam and straightforward operation, the presented source compares favorably, for smaller-scale applications, to other traditional ion source designs, typically conceived for large-scale particle accelerators or large scientific facilities, which may not be as convenient nor as cost-effective for industrial or medical applications in smaller environments.

## Acknowledgements

The authors are grateful to the Department of Economic Development and Infrastructures of the Basque Government for partial support of this work under the PIT30 Collaboration agreement with UPV/EHU, and the Elkartek project KK-2018/00020. They are also grateful to MINECO for partial support of this work under project DPI2017-82373-R. Also the financial support provided by the Departament of Education of the Basque Government, through project with ref. IT1104-16.